\documentclass[aps,pra,onecolumn,a4paper]{revtex4}%
\usepackage[left=1.25in,right=1.25in,top=1in,bottom=1in]{geometry}
\usepackage{amssymb}
\usepackage{setspace}
\usepackage{graphicx}
\usepackage{amsmath}
\usepackage{amsfonts}
\usepackage{amsthm}
\usepackage{tikz}
\usepackage{accents}%
\providecommand{\U}[1]{\protect\rule{.1in}{.1in}}

\DeclareMathOperator{\Tr}{Tr}

\selectfont{}

\newcommand{\SP}[1]{{{\textcolor{black}{#1}}}}

\begin{document}
\title{Noisy receivers for quantum illumination}
\author{Athena Karsa}
\affiliation{Department of Computer Science, University of York, York YO10 5GH, UK}
\author{Stefano Pirandola}
\affiliation{Department of Computer Science, University of York, York YO10 5GH, UK}
\date{\today}

\begin{abstract}
Quantum illumination (QI) promises unprecedented performances in target detection but there are \SP{various} problems surrounding its implementation. Where target ranging is a concern, signal and idler recombination forms a crucial barrier to the protocol's success. This could \SP{potentially} be mitigated if performing a measurement on the idler mode could still yield a quantum advantage. In this paper we investigate the QI protocol for a generically correlated Gaussian source and study the phase-conjugating (PC) receiver, deriving the associated SNR in terms of the signal and idler energies, and their cross-correlations, which may be readily adapted to incorporate added noise due to Gaussian measurements. We confirm that a heterodyne measurement performed on the idler mode leads to a performance which asymptotically approaches that of a coherent state with homodyne detection. However, if the signal mode is affected by heterodyne but the idler mode is maintained clean, the performance asymptotically approaches that of the PC receiver without any added noise.
\end{abstract}

\maketitle

\section{Introduction}
Quantum illumination (QI) \cite{pirandola2018advances,lloyd2008enhanced,tan2008quantum,zhangexp,lopaevaexp} is an entanglement-based protocol able to detect the presence of a
low-reflectivity object embedded in bright thermal noise, even in the case
where the signal employed is itself very weak. Using an optimum quantum receiver it offers a 6 dB  advantage in error probability exponent over the best possible classical strategy using the same transmitted energy. Such an advantage persists despite the fact that all entanglement is lost during the process \cite{zhangent} - lending itself to being of particular use in the microwave regime where the ambient background is inherently high \cite{barzanjehmicrowave}. 

To date, the specifics of such an optimum receiver for QI remains unknown without access to a quantum computer. There have, however, been several proposals for practical receiver designs, the best of which are the sub-optimal optical parametric amplifier (OPA) and phase-conjugating (PC) receivers \cite{guhareceiver}, achieving up to 3 dB in performance advantage. The 3 dB performance deficit is owing to the fact that these receivers operate based on Gaussian local operations which are known to be not optimal for general mixed-state discrimination \cite{calsamiglia2010local,bandyolocality}. Employing nonlinear operations, Zhuang \textit{et al.} \cite{FFSFG,quntaoNPtarget,quntaoRFtarget} used sum-frequency generation (SFG) alongside a feed-forward (FF) mechanism to show that the full QI advantage could theoretically be attained, however the physical implementation of the FF-SFG receiver is yet out of reach.

Even though low signal energy is one of the key ingredients for QI's advantage, this inherent property of microwave photons make their detection difficult such that single photon counting forms a great obstacle for any experiment in the microwave domain. This is despite the fact that this task is generally quite straightforward in other regimes with efficient optical photon counters being widely available \cite{eichler2012characterizing}. The actual measurement procedure forms a crucial and fundamental design aspect of any QI receiver, particularly in the microwave domain, with interesting progress being demonstrated by recent experiments~\cite{luong2019receiver, shabirQI}.

Further to questions regarding receiver design, idler storage poses another issue particularly with respect to target-ranging problems. In QI an entangled photon pair is created with one forming the signal and the other, the idler, stored for later joint measurement. In scenarios where the range, and thus return time of the signal, is unknown, or even a measure to be determined, idler storage forms a crucial aspect of the protocol necessary for its success. 

\SP{A potential solution is to perform a measurement on the idler photon, mitigating issues associated with its storage, and combine the result with that of the returning signal. In microwave QI, these measurement results take the form of quadrature voltages which may be used to reconstruct the annihilation operators of the modes; in turn, these may be post-processed to simulate potential receivers for QI, such as the digital PC receiver~\cite{shabirQI}. Despite the fact that the collected data can be used in this way, real-time implementation of such a strategy cannot beat the optimal performance of coherent states, as already discussed in Ref.~\cite{shabirQI} and further investigated here.}





In this paper we consider the QI protocol using a generic source modelled as a two-mode Gaussian state with arbitrary quadrature correlations.  Keeping in the domain of Gaussian linear operations, we study the PC receiver in terms of its effective signal-to-noise ratio (SNR) for our generic source. We consider various cases of added noise from, for example, the application of a heterodyne measurement on one or both of the source's modes, comparing their performances of these various receivers and determine their absolute performance capabilities relative to the optimal classical method using coherent states with homodyne detection.

\section{Basics of the quantum illumination protocol}

Consider the production of $M$ independent signal-idler mode pairs, $\{\hat
{a}_{S}^{(k)},\hat{a}_{I}^{(k)}\}$; $1\leq k\leq M$, with mean number of
photons per mode given by $N_{S}$ and $N_{I}$ for the signal and idler modes,
respectively. The signal ($S$) mode is sent out to some target region while
the idler ($I$) mode is retained at the source for later joint measurement.
Their joint state, $\hat{\rho}_{S,I}$, is modelled as a two-mode, zero mean
Gaussian state~\cite{RMP} with covariance matrix (CM) given
by~\cite{Notation}
\begin{equation}
\mathbf{V}_{S,I}=\frac{1}{2}%
\begin{pmatrix}
\nu\mathbf{1} & c\mathbf{Z}\\
c\mathbf{Z} & \mu\mathbf{1}%
\end{pmatrix}
,~\left\{
\begin{array}
[c]{l}%
\mathbf{1}:=\mathrm{diag}(1,1),\\
\mathbf{Z}:=\mathrm{diag}(1,-1),
\end{array}
\right.  \label{eq1}%
\end{equation}
where $\nu:=2N_{S}+1$, $\mu:=2N_{I}+1$ and $c$ quantifies the quadrature
correlations between the two modes such that $0\leq c\leq2\sqrt{N_{S}%
(N_{I}+1)}$. In the case where the signal-idler mode pairs are maximally
entangled we have $c=c_{q}:=2\sqrt{N_{S}(N_{I}+1)}$ while the case
$c=c_{d}:=2\sqrt{N_{S}N_{I}}$ renders the state
just-separable~\cite{EntBreak,ModiDiscord}. Recall that, for $c=c_{q}$,
the state is known as two-mode squeezed vacuum (TMSV) state~\cite{RMP}.

Under hypothesis $H_{0}$, the target is absent so that the returning mode
$\hat{a}_{R}=\hat{a}_{B}$, where $\hat{a}_{B}$ is in a thermal state with mean
number of photons per mode $N_{B}\gg1$. Under hypothesis $H_{1}$, the target
is present such that $\hat{a}_{R}=\sqrt{\kappa}\hat{a}_{S}+\sqrt{1-\kappa}%
\hat{a}_{B}$, where $\kappa\ll1$, and $\hat{a}_{B}$ is in a thermal state with
mean number of photons per mode $N_{B}/(1-\kappa)$, so that the mean noise
photon number is equal under both hypotheses (no passive signature). The conditional joint state,
$\hat{\rho}_{R,I}^{i}$ for $i=0,1$, of the returning ($R$) mode and the
retained idler is given by, under hypotheses $H_{0}$ and $H_{1}$,
respectively,
\begin{equation}
\mathbf{V}_{R,I}^{0}=\frac{1}{2}%
\begin{pmatrix}
\omega\mathbf{1} & 0\\
0 & \mu\mathbf{1}%
\end{pmatrix}
, \label{eq2}%
\end{equation}%
\begin{equation}
\mathbf{V}_{R,I}^{1}=\frac{1}{2}%
\begin{pmatrix}
\gamma\mathbf{1} & \sqrt{\kappa}c\mathbf{Z}\\
\sqrt{\kappa}c\mathbf{Z} & \mu\mathbf{1}%
\end{pmatrix}
, \label{eq3}%
\end{equation}
where we set $\omega:=2N_{B}+1$ and $\gamma:=2\kappa N_{S}+\omega$.

At this point the binary decision between target absence and presence is
reduced to the discrimination of the two quantum states $\hat{\rho}_{R,I}^{i}$ with
$i=0,1$ \cite{cheflesQSD,barnettQSD,cheflesstrategies}. The total error in such a discrimination is given by a linear combination of two error types, $P_{\mathrm{min}} = \pi_0 P(1|H_0)+ \pi_1 P(0|H_1)$, where $\pi_0$ and $\pi_1$ can be interpreted as the a priori probabilities that we assign to the occurrence of each hypothesis. For equally-likely hypotheses, the optimal measurement for the
discrimination is the dichotomic positive-operator valued measure
(POVM)~\cite{helstrom1969quantum} $E^{0}=\Pi(\gamma_{+})$, $E^{1}=1-\Pi
(\gamma_{+})$, where $\Pi(\gamma_{+})$ is the projector on the positive part
$\gamma_{+}$ of the Helstrom matrix $\gamma:= \hat{\rho}_{R,I}^{0} - \hat
{\rho}_{R,I}^{1} $. Associated with such a discrimination is the minimum error
probability given by the Helstrom bound, $P_{\mathrm{min}} = \left[
1-D(\hat{\rho}_{R,I}^{0},\hat{\rho}_{R,I}^{1}) \right]  /2$ where $D(\hat
{\rho}_{R,I}^{0},\hat{\rho}_{R,I}^{1}) := \Tr |\hat{\rho}_{R,I}^{0} -\hat
{\rho}_{R,I}^{1}|/2$ is the trace distance~\cite{watrous2018quantum}.

Due to analytical difficulty, we may instead compute bounds on the Helstrom
error probability such as the quantum Chernoff bound (QCB)~\cite{QCB}
\begin{align}
P_{\mathrm{min}}  &  \leq P_{\mathrm{QCB}}:=\frac{1}{2}\left(  \inf_{0\leq
s\leq1}C_{s}\right)  ,\nonumber\\
C_{s}  &  :=\Tr\left[  (\hat{\rho}_{R,I}^{0})^{s}(\hat{\rho}_{R,I}^{1}%
)^{1-s}\right]  , \label{QCB}%
\end{align}
where the minimisation of the $s$-overlap $C_{s}$ occurs over all $0\leq
s\leq1$. Note that, though not considered in this work, we can easily extend the quantum Chernoff bound to cover cases where the two hypotheses are not equiprobable:
\begin{equation}
P_{\mathrm{QCB}}:= \inf_{0 \leq s \leq 1} \pi_0^{s}\pi_1^{1-s} \Tr\left[  (\hat{\rho}_{R,I}^{0})^{s}(\hat{\rho}_{R,I}^{1})^{1-s}\right].
\end{equation}
For the problem under study, the minimum is achieved for $s=1/2$ that
corresponds to the simpler quantum Bhattacharyya bound~\cite{RMP}
\begin{equation}
P_{\mathrm{QBB}}:=\frac{1}{2}\Tr\left[  \sqrt{\hat{\rho}_{R,I}^{0}}\sqrt
{\hat{\rho}_{R,I}^{1}}\right]  . \label{QBB}%
\end{equation}
In particular, there is a closed analytical formula for computing $C_{s}$ for
the QCB between two arbitrary multimode Gaussian states (see
Appendix~\ref{App1}). Using this formula, we can certainly compute the QCB
between the two possible output states given in Eqs.~(\ref{eq2})
and~(\ref{eq3}), but the expression is too long to be exhibited here.

In the absence of an idler the best strategy is to use coherent states. The
signal is prepared in the coherent state $|\sqrt{N_{S}}\rangle$ which is then
sent out to some target region. Under $H_{0}$, the received returning mode is
in a thermal state with mean photon number $N_{B}$ and covariance matrix equal
to $(\omega/2)\mathbf{1}$, i.e., $\hat{a}_{R}=\hat{a}_{B}$. Under $H_{1}$, the
signal is mixed with the background such that $\hat{a}_{R}=\sqrt{\kappa}%
\hat{a}_{S}+\sqrt{1-\kappa}\hat{a}_{B}$ with $\kappa\in(0,1)$, corresponding
to a displaced thermal state with mean vector $(\sqrt{\kappa N_{S}},0)$ and
covariance matrix $(\omega/2)\mathbf{1}$. The QCB of such a coherent state
transmitter may be readily computed and takes the exact
form~\cite{tan2008quantum}
\begin{equation}
P_{\mathrm{QCB,CS}}\leq\frac{1}{2}e^{-M\kappa N_{S}\left(  \sqrt{N_{B}%
+1}-\sqrt{N_{B}}\right)  ^{2}}. \label{csQCB}%
\end{equation}
Achieving Eq.~(\ref{csQCB}) requires to use of an optimal receiver whose
structure is not known. The best practical strategy for the reception of
coherent states is homodyne detection whose measurement operators are
projectors over the quadrature basis. It is best used when the optical field
phase is maintained across the detection protocol so that each of the $M$
pulses may be coherently integrated before a binary test can be carried out on
the outcome. In such a case the false-alarm probability, $P^{\mathrm{fa}}=P(1|H_0)$,
and missed-detection probability, $P^{\mathrm{md}}=P(0|H_1)$, are given by
\begin{align}
P_{\mathrm{CS},\text{hom}}^{\text{fa}}(x)  &  =\frac{1}{2}\operatorname{erfc}%
\left(  \frac{x}{\sqrt{M(2N_{B}+1)}}\right)  ,\label{eqq1}\\
P_{\mathrm{CS},\text{hom}}^{\text{md}}(x)  &  =\frac{1}{2}\operatorname{erfc}%
\left(  \frac{M\sqrt{2\kappa N_{S}}-x}{\sqrt{M(2N_{B}+1)}}\right)  ,
\label{eqq2}%
\end{align}
where $\operatorname{erfc}(z):=1-2\pi^{-1/2}\int_{0}^{z}\exp(-t^{2})dt$ is the
complementary error function. For equally-likely hypotheses, Eqs.~(\ref{eqq1})
and~(\ref{eqq2}) may be combined and minimised over $x$ to give the minimum
average error probability for homodyne detection and coherent integration
\begin{equation}
P_{\mathrm{CS},\mathrm{hom}}=\frac{P_{\mathrm{CS},\mathrm{hom}}^{\mathrm{fa}%
}+P_{\mathrm{CS},\mathrm{hom}}^{\mathrm{md}}}{2}=\frac{1}{2}%
\operatorname{erfc}\left(  \sqrt{\frac{M\kappa N_{S}}{4N_{B}+2}}\right)  .
\label{homodyne}%
\end{equation}

\section{The phase-conjugating receiver}

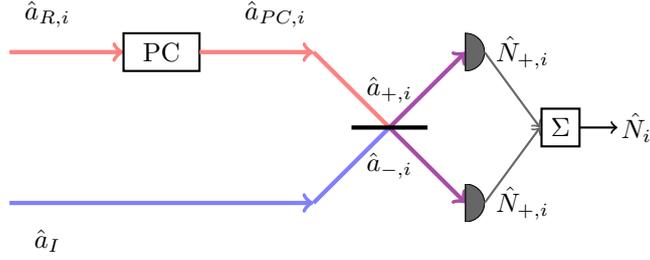
\begin{figure}[ptb]
\centering
\begin{tikzpicture}
\node at (0.5,3.5) {$\hat{a}_{R,i}$};
\draw [->,red!50,ultra thick] (0,3) -- (1.5,3);
\node at (2,3) {PC};
\draw[thick] (1.5,2.75) rectangle (2.5,3.25);
\node at (3.5,3.5) {$\hat{a}_{PC,i}$};
\draw [->,red!50,ultra thick] (2.5,3) -- (4,3);
\node at (0.5,0.5) {$\hat{a}_{I}$};
\draw [->,blue!50,ultra thick] (0,1) -- (4,1);
\draw [-,red!50,ultra thick] (4,3) -- (5,2);
\draw [->,violet!70,ultra thick] (5,2) -- (6,1);
\draw [-,blue!50,ultra thick] (4,1) -- (5,2);
\draw [->,violet!70,ultra thick] (5,2) -- (6,3);
\draw [-,black,ultra thick] (4.5,2) -- (5.5,2);
\node at (5,2.5) {$\hat{a}_{+,i}$};
\node at (5,1.5) {$\hat{a}_{-,i}$};
\begin{scope}
\clip (6,2.75) rectangle (6.25,3.25);
\draw[fill=black!60] (6,3) circle(0.25);
\draw (6,2.75) -- (6,3.25);
\end{scope}
\begin{scope}
\clip (6,0.75) rectangle (6.25,1.25);
\draw[fill=black!60] (6,1) circle(0.25);
\draw (6,0.75) -- (6,1.25);
\end{scope}
\draw[->,black!60,thick] (6.25,3)--(6.98,2);
\node at (6.75,3) {$\hat{N}_{+,i}$};
\draw[->,black!60,thick] (6.25,1)--(6.98,2);
\node at (6.75,1) {$\hat{N}_{+,i}$};
\node at (7.25,2) {$\Sigma$};
\draw[thick] (7,1.75) rectangle (7.5,2.25);
\draw[->,thick] (7.5,2) -- (8,2);
\node at (8.25,2) {$\hat{N}_i$};
\end{tikzpicture}
\caption{The phase-conjugating (PC) receiver used to calculate the SNR of the
QI protocol. Each of the $M$ copies of the returning signal modes are
phase-conjugated before being mixed with each of the individual corresponding
retained idler modes in a 50-50 beamsplitter. These outputs are photodetected
with the difference between the two detectors' outputs corresponding to an
outcome equivalent to that of the total photon number operator. This is used
as input to a threshold detector which makes the binary decision: target
absent or target present.}%
\label{PCreceiverdesign}%
\end{figure}

The phase-conjugating (PC) receiver~\cite{guhareceiver} is one possible
practical detector for QI. As depicted in
Fig.~\ref{PCreceiverdesign}, this receiver phase-conjugates all $M$ returning
modes $\hat{a}_{R,i}^{(k)}$, where $1\leq k\leq M$ and $i=0,1$ (corresponding
to the two hypotheses $H_{0}$\ and $H_{1}$), according to%
\begin{equation}
\hat{a}_{PC,i}=\sqrt{2}\hat{a}_{v}+\hat{a}_{R,i}^{\dag}, \label{eq4}%
\end{equation}
where $\hat{a}_{v}$ is the vacuum operator. Since the creation and
annihilation operators are defined in terms of quadratures, $\hat{q}$ and
$\hat{p}$, via $\hat{a}=(\hat{q}+i\hat{p})/\sqrt{2}$ and $\hat{a}^{\dag}%
=(\hat{q}-i\hat{p})/\sqrt{2}$, respectively, we may recast Eq.~(\ref{eq4}) for
$\hat{X}=(\hat{q},\hat{p})^{T}$. Thus the PC receiver transforms quadratures
as
\begin{equation}
\hat{X}_{PC,i}=\sqrt{2}\hat{X}_{v}+\mathbf{Z}\hat{X}_{R,i}. \label{eq5}%
\end{equation}
One can write a corresponding action on the conditional covariance matrices
$\mathbf{V}_{R,I}^{i}$ in Eqs.~(\ref{eq2}) and~(\ref{eq3}) (see
Appendix~\ref{App2} for detailed calculations). The individual phase-conjugated
signal modes are then mixed with the corresponding retained idler modes on a
50-50 beamsplitter whose outputs are given by
\begin{equation}
\hat{a}_{\pm,i}=\frac{\hat{a}_{PC,i}\pm\hat{a}_{I}}{\sqrt{2}}, \label{eq8}%
\end{equation}
or, equivalently,
\begin{equation}
\hat{X}_{\pm,i}=\frac{\hat{X}_{PC,i}\pm\hat{X}_{I}}{\sqrt{2}}. \label{eq9}%
\end{equation}
It is these modes with output conditional covariance matrices $\mathbf{V}%
_{\pm}^{i}$, for $i=0,1$, which are then photodetected yielding photon counts
equivalent to measurement outcomes of the number operator $\hat{N}_{\pm
,i}=\hat{a}_{\pm,i}^{\dag}\hat{a}_{\pm,i}$.

The binary decision is made by computing the difference between the two
detectors' outputs~\cite{barzanjehmicrowave}, equivalent to the measurement
outcome of the operator
\begin{equation}
\hat{N}_{i}=\hat{N}_{+,i}-\hat{N}_{-,i}. \label{eq15}%
\end{equation}
Since the QI protocol uses a very large number of copies $M$ of the
signal-idler mode pairs, the central limit theorem applies to our
measurements. That is, the measurement $\hat{N}_{i}$ yields a
Gaussian-distributed random variable, conditioned on the hypothesis. Thus we
can write the QI receiver's signal-to-noise ratio (SNR), for hypotheses
with equal prior probabilities, as~\cite{guhareceiver}
\begin{equation}
\mathrm{SNR}=\frac{\left(  \langle\hat{N}_{1}\rangle-\langle\hat{N}_{0}%
\rangle\right)  ^{2}}{2\left(  \sqrt{\langle\Delta\hat{N}_{1}^{2}\rangle
}+\sqrt{\langle\Delta\hat{N}_{0}^{2}\rangle}\right)  ^{2}}, \label{eq16}%
\end{equation}
where $\langle\hat{O}_{i}\rangle$ and $\langle\Delta\hat{O}_{i}^{2}\rangle$,
for $i=0,1$, are the conditional means and variances of measurement $\hat
{O}_{i}$, respectively, and the notation $\langle\dots\rangle$ denotes an
average over all $M$ copies.

In Appendix~\ref{App2} we explicitly calculate each of these quantities in turn
and find the single-mode SNR, Eq.~(\ref{eq16}) of the PC receiver for a
generic two-mode Gaussian state source as in Eq.~(\ref{eq1}) with quadrature
correlations $c$. This SNR is given by
\begin{equation}
\mathrm{SNR}_{\mathrm{PC}}=\frac{\kappa c^{2}}{\left(  \sqrt{\kappa c^{2}%
+\mu(1+\gamma)}+\sqrt{\mu(1+\omega)}\right)  ^{2}}. \label{eq25}%
\end{equation}
This directly relates to its error probability after $M$ uses, for
equally-likely hypotheses, satisfying~\cite{guhareceiver}
\begin{equation}
P_{\mathrm{PC}}^{(M)}=\frac{1}{2}\mathrm{erfc}\left(  \sqrt{M\mathrm{SNR}%
_{\mathrm{PC}}}\right)  . \label{receiver}%
\end{equation}

\section{Comparison between receivers with added noise}

It is easy to modify the final formula in Eq.~(\ref{eq25}) to include the
presence of extra noise on the idler mode ($\hat{a}_{I}$) and returning signal
mode ($\hat{a}_{R,i}$) \textit{before} the action of the PC receiver. Assuming
that this extra noise is Gaussian added noise with variances $\varepsilon_{I}$
(for the idler) and $\varepsilon_{R}$ (for the returning signal), we may write
the same SNR\ in Eq.~(\ref{eq25}) up to the following replacement%
\begin{equation}
\mu\rightarrow\mu^{\prime}=\mu+\varepsilon_{I},
\end{equation}
and
\begin{equation}%
\begin{array}
[c]{c}%
\omega\rightarrow\omega^{\prime}=\omega+\varepsilon_{R},~\text{under }%
H_{0}\text{,}\\
\gamma\rightarrow\gamma^{\prime}=\gamma+\varepsilon_{R},~\text{under }%
H_{1}\text{.}%
\end{array}
\end{equation}
Let us assume that this added noise is the same amount you would get from the
application of a heterodyne measurement, so that $\varepsilon_{I(R)}=1$.
Besides the standard configuration of an entangled TMSV source and the
PC\ receiver that we denote QI+PC, consider the case where both idler and
returning signal modes are affected by the extra noise $\varepsilon
_{I}=\varepsilon_{R}=1$ before the PC receiver, a configuration that we denote QI+Het+PC. Then, consider the hybrid case where only the returning signal
is affected while the idler is noiseless or \textquotedblleft
calibrated\textquotedblright, i.e., $\varepsilon_{R}=1$ and $\varepsilon
_{I}=0$, that we denote QI+Cal+PC.

Let us also consider another scenario. For the case where both idler and
returning signal modes have added noise ($\varepsilon_{I}=\varepsilon_{R}=1$),
let us assume this is indeed the effect of heterodyne detections. Let us now
assume that the outcomes are processed in the optimal way so that we may apply the classical Chernoff bound (CCB)\cite{QCB}. Recall that, for two probability
distributions, $p_{0}(i)$ and $p_{1}(i)$, the CCB is given by
\begin{equation}
\xi_{\mathrm{CCB}}=-\log\left(  \min_{0\leq s\leq1}\sum_{i}p_{0}(i)^{s}%
p_{1}(i)^{1-s}\right)  .
\end{equation}
The outcomes of the heterodyne detections are distributed according to
Gaussian probability densities that are directly related to the Wigner
functions of the states. In fact, we have
\begin{equation}
\xi_{\mathrm{CCB}}=\pi^{2}\int d^{4}\mathbf{x}\,\,W_{\mathbf{V}^{0}%
+\mathbf{1}}^{s}(\mathbf{x})W_{\mathbf{V}^{1}+\mathbf{1}}^{1-s}(\mathbf{x}).
\end{equation}
where we have the two modes' quadrature components $\mathbf{x}:=(q_{R},p_{R},q_{I},p_{I})^{T}$, $W_{\mathbf{V}^{0}%
}(\mathbf{x})$ is the Wigner function of $\hat{\rho}_{R,I}^{0}$ and
$W_{\mathbf{V}^{1}}(\mathbf{x})$ is that of $\hat{\rho}_{R,I}^{1}$. Here we
have
\begin{equation}
W_{\mathbf{V}^{i}}(\mathbf{x})=\frac{\exp\left[  -\frac{1}{2}\mathbf{x}%
^{T}(\mathbf{V}^{i})^{-1}\mathbf{x}\right]  }{4\pi^{2}\sqrt{\det\mathbf{V}%
^{i}}},
\end{equation}
where $\mathbf{V}^{i}=\mathbf{V}_{R,I}^{i}$ is given in Eqs.~(\ref{eq2})
and~(\ref{eq3}) for $i=0,1$. Denoting this case by QI+Het+CCB, we find
that
\begin{equation}
\xi_{\mathrm{QI+Het+CCB}}=\frac{4(1+N_{B})}{4+4N_{B}+\kappa N_{S}}.
\end{equation}
\begin{figure}[ptb]
\centering
\includegraphics[width=8.6cm]{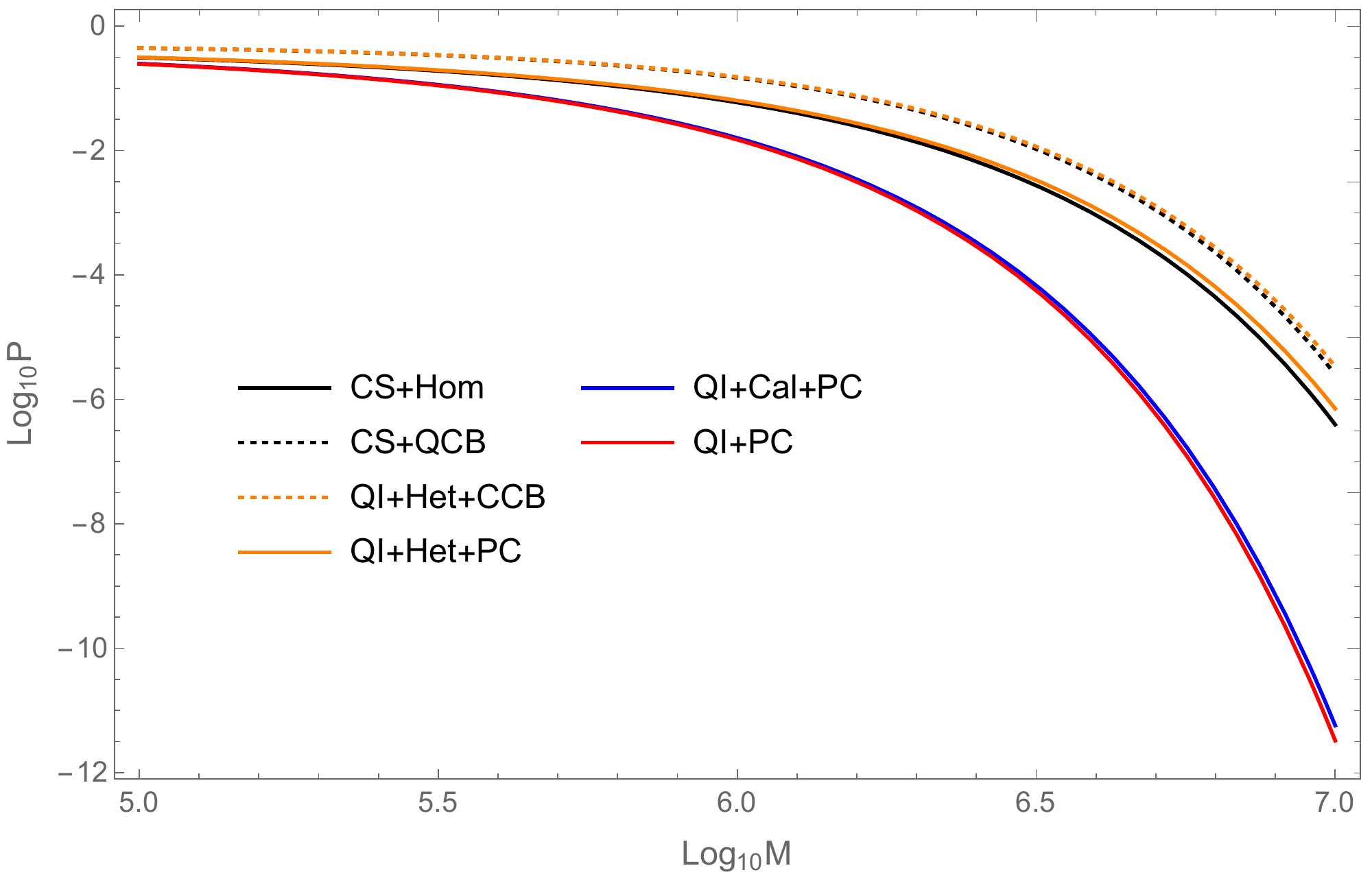}\caption{Performance
comparison of the various receivers in terms of error exponent versus
(logarithmic) number of uses $M$. The results are computed for parameter
values $N_{S}=N_{I}=1/100$,\ $c=2\sqrt{N_{S}(N_{I}+1)}$, $N_{B}=20$ and
$\kappa=1/100$.}%
\label{results1}%
\end{figure}

We compare the performances of these receivers to that given by the
coherent state QCB (CS-QCB) and coherent state transmitter with homodyne
detection (CS+Hom), given by Eqs.~(\ref{csQCB}) and~(\ref{homodyne}),
respectively. Results are shown in Fig.~\ref{results1} where we plot the error
probability exponent as a function of $M$. We see that QI+Het+PC is
outperformed by CS+Hom. Likewise, QI+Het+CCB does not surpass CS-QCB. However,
it can be seen that the hybrid case QI+Cal+PC approaches the noiseless
receiver QI+PC under optimal conditions: maximal entanglement, i.e.,
$c=2\sqrt{N_{S}(N_{I}+1)}$, symmetric low-brightness $N_{S}=N_{I}\ll1$, and
large background $N_{B}\gg1$. Performing an asymptotic expansion in the regime
of large $N_{B}$, we find that
\begin{equation}
\mathrm{SNR}_{\mathrm{QI+Cal+PC}}\rightarrow\mathrm{SNR}_{\mathrm{QI+PC}%
}=\frac{(1+N_{I})\kappa N_{S}}{2N_{B}(1+2N_{I})},
\end{equation}
and
\begin{equation}
\mathrm{SNR}_{\mathrm{QI+Het+PC}}\rightarrow\mathrm{SNR}_{\mathrm{CS+Hom}%
}=\frac{\kappa N_{S}}{4N_{B}}.
\end{equation}
The maximal advantage of QI+PC\ over CS+Hom is given by
\begin{equation}
\frac{\mathrm{SNR}_{\mathrm{QI+PC}}}{\mathrm{SNR}_{\mathrm{CS+Hom}}}%
=\frac{2(1+N_{I})}{1+2N_{I}}\rightarrow2\text{~for }N_{I}\ll1\text{.}%
\end{equation}

Our analysis above clearly shows that, whenever the idler mode is affected by an additive Gaussian 
noise that is equivalent to a heterodyne detection, the performance of coherent state transmitters cannot
be beaten. There is indeed another argument to understand why this is the case. Performing a Gaussian measurement on the idler mode of a two-mode Gaussian state remotely prepares an ensemble of Gaussian states on the signal mode~\cite{pirandola2014optimality}. In particular, if the Gaussian state is a TMSV state and the idler mode is heterodyned, then the signal mode is projected onto an ensemble of coherent states, whose average state is thermal with mean number of photons equal to the signal energy of the TMSV state.

\section{Conclusion}

In this paper we have investigated the QI protocol for a generically correlated Gaussian source, considering various receiver types. Keeping within the realms of Gaussian operations, we have paid attention to the PC receiver and studied its performance in various cases of added noise due to, for example, the action of a heterodyne measurement on one or both of the modes. The potential of performing a measurement on the idler, still retaining a quantum advantage, would mitigate one of the major problems associated with QI implementation: idler storage and later recombination with the returning signal. This is of particular concern when the problem involves target ranging, \SP{where alternative strategies should be considered~\cite{QTSP2020}.}

Under these considerations, we have modelled the PC receiver for our generic source and have derived the associated SNR in terms of the signal and idler energies, and their cross-correlations. Our SNR may be readily adapted to include additional noise associated with Gaussian measurements. Our results confirm that a heterodyne measurement performed on the idler mode leads to a performance which asymptotically approaches that of a coherent state with homodyne detection, not surpassing it. 
Interestingly, if the signal mode is affected by heterodyne but the idler mode is maintained clean, the performance asymptotically approaches that of the PC receiver without any added noise. \SP{Finally, let us mention that it would be interesting to investigate these aspects within the setting of unambiguous quantum discrimination~\cite{Janos2005}.}

\appendix

\section{Quantum Chernoff bound for multimode Gaussian states\label{App1}}

Consider two arbitrary $N$-mode Gaussian states, $\hat{\rho}_{0}%
(\mathbf{x}_{0},\mathbf{V}_{0})$ and $\hat{\rho}_{1}(\mathbf{x}_{1}%
,\mathbf{V}_{1})$, with mean $\mathbf{x}_{i}$ and CM $\mathbf{V}_{i}$ with
quadratures $\mathbf{\hat{x}}=\left(  \hat{q}_{1},\hat{p}_{1},\dots,\hat
{q}_{N},\hat{p}_{N}\right)  ^{T}$ and associated symplectic form
\begin{equation}
\mathbf{\Omega}=\bigoplus_{k=1}^{N}%
\begin{pmatrix}
0 & 1\\
-1 & 0
\end{pmatrix}
.
\end{equation}
We can write the $s$-overlap as~\cite{pirandola2008computable}%
\begin{equation}
C_{s}=2^{N}\sqrt{\frac{\det\mathbf{\Pi}_{s}}{\det\mathbf{\Sigma}_{s}}}%
\exp\left(  -\frac{\mathbf{d}^{T}\mathbf{\Sigma}_{s}^{-1}\mathbf{d}}%
{2}\right)  ,
\end{equation}
where $\mathbf{d}=\mathbf{x}_{0}-\mathbf{x}_{1}$. Here $\mathbf{\Pi}_{s}$ and
$\mathbf{\Sigma}_{s}$ are defined as
\begin{equation}
\mathbf{\Pi}_{s}:=G_{s}(\mathbf{V}_{0}^{\oplus})G_{1-s}(\mathbf{V}_{1}%
^{\oplus}),
\end{equation}
\vspace{-0.5cm}
\begin{equation}
\mathbf{\Sigma}_{s}:=\mathbf{S}_{0}\left[  \Lambda_{s}\left(  \mathbf{V}%
_{0}^{\oplus}\right)  \right]  \mathbf{S}_{0}^{T}+\mathbf{S}_{1}\left[
\Lambda_{1-s}\left(  \mathbf{V}_{1}^{\oplus}\right)  \right]  \mathbf{S}%
_{1}^{T},
\end{equation}
introducing the two real functions
\begin{align}
G_{s}(x)  &  =\frac{1}{(x+1/2)^{s}-(x-1/2)^{s}}\nonumber\\
\Lambda_{s}(x)  &  =\frac{(x+1/2)^{s}+(x-1/2)^{s}}{(x+1/2)^{s}-(x-1/2)^{s}},
\end{align}
calculated over the Williamson forms $\mathbf{V}_{i}^{\oplus}%
:=\mathbf{\bigoplus}_{k=1}^{N}\nu_{i}^{k}\mathbf{1}_{2}$, where $\mathbf{V}%
_{i}^{\oplus}\mathbf{=S}_{i}\mathbf{\mathbf{V}}_{i}^{\oplus}\mathbf{S}_{i}%
^{T}$ for symplectic $\mathbf{S}_{i}$\ and $\nu_{i}^{k}\geq1/2$ are the
symplectic spectra~\cite{serafini2003symplectic,pirandola2009correlation}.

\section{SNR for the phase-conjugating receiver\label{App2}}

The phase-conjugating (PC) receiver~\cite{guhareceiver}, see
Fig.~\ref{PCreceiverdesign} in the main text, phase-conjugates all $M$
returning modes $\hat{a}_{R,i}^{(k)}$, where $1\leq k\leq M$ and $i=0,1$
(corresponding to the two hypotheses $H_{0}$\ and $H_{1}$), according to%
\begin{equation}
\hat{a}_{PC,i}=\sqrt{2}\hat{a}_{v}+\hat{a}_{R,i}^{\dag}, \label{eqa4}%
\end{equation}
where $\hat{a}_{v}$ is the vacuum operator. Since the creation and
annihilation operators are defined in terms of quadratures, $\hat{q}$ and
$\hat{p}$, via $\hat{a}=(\hat{q}+i\hat{p})/\sqrt{2}$ and $\hat{a}^{\dag}%
=(\hat{q}-i\hat{p})/\sqrt{2}$, respectively, we may recast Eq.~(\ref{eqa4})
for $\hat{X}=(\hat{q},\hat{p})^{T}$. Thus the PC receiver transforms
quadratures as
\begin{equation}
\hat{X}_{PC,i}=\sqrt{2}\hat{X}_{v}+\mathbf{Z}\hat{X}_{R,i}, \label{eqa5}%
\end{equation}
and the corresponding conditional covariance matrices of the return-idler
states are given by
\begin{equation}
\mathbf{V}_{PC,I}^{0}=\frac{1}{2}%
\begin{pmatrix}
(\omega+1)\mathbf{1} & 0\\
0 & \mu\mathbf{1}%
\end{pmatrix}
, \label{eqb6}%
\end{equation}%
\begin{equation}
\mathbf{V}_{PC,I}^{1}=\frac{1}{2}%
\begin{pmatrix}
(\gamma+1)\mathbf{1} & \sqrt{\kappa}c\mathbf{Z}\\
\sqrt{\kappa}c\mathbf{Z} & \mu\mathbf{1}%
\end{pmatrix}
. \label{eqb7}%
\end{equation}
The individual phase-conjugated signal modes are then mixed with the
corresponding retained idler modes on a 50-50 beamsplitter whose outputs are
given by
\begin{equation}
\hat{a}_{\pm,i}=\frac{\hat{a}_{PC,i}\pm\hat{a}_{I}}{\sqrt{2}}, \label{eqb8}%
\end{equation}
or, equivalently,
\begin{equation}
\hat{X}_{\pm,i}=\frac{\hat{X}_{PC,i}\pm\hat{X}_{I}}{\sqrt{2}}. \label{eqb9}%
\end{equation}
We construct the output conditional covariance matrices $\mathbf{V}_{\pm}^{i}%
$, for $i=0,1$, by considering the individual components, e.g., for $H_{0}$:
\begin{align}
\langle\hat{X}_{+,0}^{2}\rangle &  =\frac{1}{2}\left[  \langle\hat{X}%
_{PC,0}^{2}\rangle+\langle\hat{X}_{0}^{2}\rangle+2\langle\hat{X}_{PC,0}\hat
{X}_{0}\rangle\right] \nonumber\\
&  =\frac{1}{2}\left[  \frac{1}{2}(\omega+1)\mathbf{1}+\frac{1}{2}%
\mu\mathbf{1}\right] \nonumber\\
&  =\frac{1}{2}\left(  \frac{\omega+1+\mu}{2}\right)  \mathbf{1}=\langle
\hat{X}_{-,0}^{2}\rangle\label{eqb10}\\
\langle\hat{X}_{+,0}\hat{X}_{-,0}\rangle &  =\frac{1}{2}\left[  \langle\hat
{X}_{PC,0}^{2}\rangle-\langle\hat{X}_{0}^{2}\rangle\right] \nonumber\\
&  =\frac{1}{2}\left[  \frac{1}{2}(\omega+1)\mathbf{1}-\frac{1}{2}%
\mu\mathbf{1}\right] \nonumber\\
&  =\frac{1}{2}\left(  \frac{\omega+1-\mu}{2}\right)  \mathbf{1},
\label{eqb12}%
\end{align}
and similarly for $H_{1}$:
\begin{align}
\langle\hat{X}_{+,1}^{2}\rangle &  =\frac{1}{2}\left[  \langle\hat{X}%
_{PC,1}^{2}\rangle+\langle\hat{X}_{I}^{2}\rangle+2\langle\hat{X}_{PC,1}\hat
{X}_{I}\rangle\right] \nonumber\\
&  =\frac{1}{2}\left[  \frac{1}{2}(\gamma+1)\mathbf{1}+\frac{1}{2}%
\mu\mathbf{1}+\sqrt{\kappa}c\mathbf{1}\right] \nonumber\\
&  =\frac{1}{2}\left(  \frac{\gamma+1+\mu}{2}+\sqrt{\kappa}c\right)
\mathbf{1},\label{eqb10b}\\
\langle\hat{X}_{-,1}^{2}\rangle &  =\frac{1}{2}\left[  \langle\hat{X}%
_{PC,1}^{2}\rangle+\langle\hat{X}_{I}^{2}\rangle-2\langle\hat{X}_{PC,1}\hat
{X}_{I}\rangle\right] \nonumber\\
&  =\frac{1}{2}\left(  \frac{\gamma+1+\mu}{2}-\sqrt{\kappa}c\right)
\mathbf{1},\label{eqb11}\\
\langle\hat{X}_{+,1}\hat{X}_{-,1}\rangle &  =\frac{1}{2}\left[  \langle\hat
{X}_{PC,1}^{2}\rangle-\langle\hat{X}_{I}^{2}\rangle\right] \nonumber\\
&  =\frac{1}{2}\left(  \frac{\gamma+1-\mu}{2}\right)  \mathbf{1}.
\label{eqb12b}%
\end{align}

Thus the output conditional covariance matrices $\mathbf{V}_{\pm}^{i}$ are
given by
\begin{equation}
\mathbf{V}_{\pm}^{0} =
\begin{pmatrix}
\alpha_{+} \mathbf{1} & \alpha_{-} \mathbf{1}\\
\alpha_{-} \mathbf{1} & \alpha_{+} \mathbf{1}%
\end{pmatrix}
, \label{eqb13}%
\end{equation}
\begin{equation}
\mathbf{V}_{\pm}^{1} =
\begin{pmatrix}
\beta_{+} \mathbf{1} & \gamma^{*} \mathbf{1}\\
\gamma^{*} \mathbf{1} & \beta_{-} \mathbf{1}%
\end{pmatrix}
, \label{eqb14}%
\end{equation}
where $\alpha_{\pm} = (\omega+1 \pm\mu)/4$, $\beta_{\pm} = (\gamma+ 1 + \mu
\pm2\sqrt{\kappa} c)/4$ and $\gamma^{*}=(\gamma+1 - \mu)/4$. It is these modes
which are then photodetected yielding photon counts equivalent to measurement
outcomes of the number operator $\hat{N}_{\pm,i} = \hat{a}_{\pm,i}^{\dag}
\hat{a}_{\pm,i}$. The binary decision is made by computing the difference
between the two detectors' outputs~\cite{barzanjehmicrowave}, equivalent to
the measurement outcome of the operator
\begin{equation}
\hat{N}_{i} = \hat{N}_{+,i}-\hat{N}_{-,i}. \label{eqb15}%
\end{equation}
Since the QI protocol uses a very large number of copies, $M$, of the
signal-idler mode pairs the central limit theorem applies to our measurements.
That is, the measurement $\hat{N}_{i}$ yields a Gaussian-distributed random
variable, conditioned on the hypothesis. Thus we can write that the QI
receiver's signal-to-noise ratio (SNR), for hypotheses with equal prior
probabilities, satisfies~\cite{guhareceiver}
\begin{equation}
\mathrm{SNR}=\frac{\left(  \langle\hat{N}_{1} \rangle- \langle\hat{N}_{0}
\rangle\right)  ^{2}}{2 \left(  \sqrt{\langle\Delta\hat{N}_{1}^{2} \rangle} +
\sqrt{\langle\Delta\hat{N}_{0}^{2} \rangle} \right)  ^{2}}, \label{eqb16}%
\end{equation}
where $\langle\hat{O}_{i} \rangle$ and $\langle\Delta\hat{O}_{i}^{2} \rangle$,
for $i=0,1$, are the conditional means and variances of measurement $\hat
{O}_{i}$, respectively, and the notation $\langle\dots\rangle$ denotes an
average over all $M$ copies.

To evaluate the PC receiver's SNR for the QI protocol we begin by considering
the number operator in terms of quadrature operators, $\hat{N} = \hat{a}%
^{\dag} \hat{a} := (\hat{q}^{2} + \hat{p}^{2} -1)/2$. Thus we can write the
mean number of photons as
\begin{equation}
\langle\hat{N} \rangle= \frac{\langle\hat{q}^{2} \rangle+ \langle\hat{p}^{2}
\rangle- 1}{2}. \label{eqb17}%
\end{equation}
Applying this to the conditional covariance matrices given by
Eqs.~(\ref{eqb13}) and~(\ref{eqb14}), we can compute the numerator of the SNR,
Eq.~(\ref{eqb16}) for the QI PC receiver as
\begin{align}
\left(  \langle\hat{N}_{1} \rangle- \langle\hat{N}_{0} \rangle\right)  ^{2}
&  = \left(  \langle\hat{N}_{+,1} \rangle- \langle\hat{N}_{-,1} \rangle-
\langle\hat{N}_{+,0} \rangle+ \langle\hat{N}_{-,0} \rangle\right)
^{2}\nonumber\\
&  =\left(  \beta_{+} - \beta_{-} \right)  ^{2}\nonumber\\
&  = \kappa c^{2}.
\end{align}
Considering the photon number variance, we have that
\begin{align}
\langle\Delta\hat{N}^{2} \rangle &  := \langle\hat{N}^{2} \rangle- \langle
\hat{N} \rangle^{2}\nonumber\\
&  = \langle( \hat{N}_{+} - \hat{N}_{-})^{2} \rangle- \langle\hat{N}_{+} -
\hat{N}_{-}\rangle^{2}\nonumber\\
&  = \langle\Delta\hat{N}_{+}^{2} \rangle+ \langle\Delta\hat{N}_{-}^{2}
\rangle+\underbrace{ 2 \left[  \langle\hat{N}_{+} \rangle\langle\hat{N}_{-}
\rangle- \langle\hat{N}_{+} \hat{N}_{-} \rangle\right]  }_{(*)}. \label{eqb18}%
\end{align}
For the first two terms we begin by considering the form of the photon number
variance in terms of quadrature operators using Eq.~(\ref{eqb17})
\begin{align}
\langle\Delta\hat{N}_{\pm}^{2} \rangle &  := \langle\hat{N}_{\pm}^{2} \rangle-
\langle\hat{N}_{\pm} \rangle^{2}\nonumber\\
&  = \frac{1}{4} (\langle\hat{q}_{\pm}^{4} \rangle- \langle\hat{q}_{\pm}^{2}
\rangle^{2} + \langle\hat{p}_{\pm}^{4} \rangle- \langle\hat{p}_{\pm}^{2}
\rangle^{2} )\nonumber\\
&  = \frac{1}{2} (\langle\hat{q}_{\pm}^{2} \rangle^{2} + \langle\hat{p}_{\pm
}^{2} \rangle^{2}), \label{eqbvariance}%
\end{align}
where we have used the following identity for higher order Gaussian moments
\begin{equation}
\langle\hat{O}^{n} \rangle= \sigma^{n} (n-1)!!,
\end{equation}
where $\sigma= \sqrt{\langle\hat{O}^{2} \rangle}$ is the standard deviation
and $!!$ denotes the double factorial.

Rewriting the final term $(*)$ in terms of quadrature components,
Eq.~(\ref{eqb17}), we find that
\begin{align}
(*)  &  = \frac{1}{2} ( \langle\hat{q}_{+}^{2} \rangle\langle\hat{q}_{-}^{2}
\rangle- \langle\hat{q}_{+}^{2} \hat{q}_{-}^{2} \rangle+ \langle\hat{q}%
_{+}^{2} \rangle\langle\hat{p}_{-}^{2} \rangle- \langle\hat{q}_{+}^{2} \hat
{p}_{-}^{2} \rangle\nonumber\\
&  \quad+ \langle\hat{p}_{+}^{2} \rangle\langle\hat{p}_{-}^{2} \rangle-
\langle\hat{p}_{+}^{2} \hat{p}_{-}^{2} \rangle+ \langle\hat{q}_{-}^{2}
\rangle\langle\hat{p}_{+}^{2} \rangle- \langle\hat{q}_{-}^{2} \hat{p}_{+}^{2}
\rangle)\nonumber\\
&  =-(\langle\hat{q}_{+} \hat{q}_{-} \rangle^{2} + \langle\hat{q}_{+} \hat
{p}_{-} \rangle^{2} + \langle\hat{p}_{+} \hat{p}_{-} \rangle^{2} + \langle
\hat{q}_{+} \hat{p}_{+} \rangle^{2}), \label{eqbvariance1}%
\end{align}
where we have used the following identity for multivariate higher-order
Gaussian moments,
\begin{equation}
\langle\hat{O}_{i}^{2} \hat{O}_{j}^{2} \rangle= \langle\hat{O}_{ii}^{2}
\rangle\langle\hat{O}_{jj}^{2} \rangle+ 2 \langle\hat{O}_{i} \hat{O}_{j}
\rangle^{2},
\end{equation}
where $\langle\hat{O}_{i} \hat{O}_{j} \rangle$ denotes the covariance of
Gaussian variables $\hat{O}_{i}$ and $\hat{O}_{j}$.

Computing the required variances and covariances we find
\begin{align}
&  \langle\hat{q}_{+,0}^{2} \rangle^{2} = \langle\hat{q}_{-,0}^{2} \rangle^{2}
= \langle\hat{p}_{+,0}^{2} \rangle^{2} = \langle\hat{p}_{-,0}^{2} \rangle^{2}
= \alpha_{+}^{2},\nonumber\\
&  \langle\hat{q}_{+,0} \hat{q}_{-,0} \rangle^{2} = \langle\hat{p}_{+,0}
\hat{p}_{-,0} \rangle^{2} = \alpha_{-}^{2}\nonumber\\
&  \langle\hat{q}_{+,1}^{2} \rangle^{2} = \langle\hat{p}_{+,1}^{2} \rangle^{2}
= \beta_{+}^{2},\nonumber\\
&  \langle\hat{q}_{-,1}^{2} \rangle^{2} = \langle\hat{p}_{-,1}^{2} \rangle^{2}
= \beta_{-}^{2},\nonumber\\
&  \langle\hat{q}_{+,1} \hat{q}_{-,1} \rangle^{2} = \langle\hat{p}_{+,1}
\hat{p}_{-,1} \rangle^{2} = \gamma^{*2}. \label{eqb22}%
\end{align}
Inserting these into Eqs.~(\ref{eqb18}),~(\ref{eqbvariance})
and~(\ref{eqbvariance1}), we obtain the photon number variances,
\begin{align}
\langle\Delta\hat{N}_{0}^{2} \rangle &  = \frac{1}{2} \left(  \mu(1+ \omega)
\right)  ,\\
\langle\Delta\hat{N}_{1}^{2} \rangle &  = \frac{1}{2} \left(  \kappa c^{2} +
\mu(1+ \gamma)\right)  .
\end{align}
Finally, we find that the single-mode SNR, Eq.~(\ref{eqb16}) of the PC
receiver is given by
\begin{equation}
\mathrm{SNR}_{\mathrm{PC}} = \frac{\kappa c^{2}}{ \left(  \sqrt{\kappa c^{2} +
\mu(1+\gamma)} + \sqrt{\mu(1+\omega)} \right)  ^{2}}, \label{eqb25}%
\end{equation}
which directly relates to its error probability after $M$ uses, for
equally-likely hypotheses, satisfying~\cite{guhareceiver}
\begin{equation}
P^{(M)}_{\mathrm{PC}} = \frac{1}{2} \mathrm{erfc}\left(  \sqrt{M
\mathrm{SNR}_{\mathrm{PC}}} \right)  . \label{breceiver}%
\end{equation}


\begin{thebibliography}{99}
\bibitem {Notation}Note that we are here assuming that the variance of the
vacuum noise is equal to $1/2$.
\end{thebibliography}
\end{document}